\documentclass{article}
\usepackage{graphicx} 
\usepackage[backend=biber, sorting=none]{biblatex}
\usepackage{listings}
\usepackage{hyperref}
\usepackage{authblk}

\title{Utilizing SciPy and other open source packages to provide a powerful API for materials manipulation in the Schr\"odinger Materials Suite}
\author[1,2]{Alexandr Fonari}
\author[1]{Farshad Fallah}
\author[1]{Michael Rauch}
\affil[1]{Schr\"odinger Inc., 1540 Broadway, 24th Floor. New York, NY 10036, US.}
\affil[2]{alexandr.fonari@schrodinger.com}
\providecommand{\keywords}[1]{\textbf{\textit{Keywords:}} #1}
\date{}

\begin{document}

\maketitle

\begin{abstract}
The use of several open source scientific packages in the Schr\"odinger Materials Science Suite will be discussed.
A typical workflow for materials discovery will be described, discussing how open source packages have been incorporated at every stage.
Some recent implementations of machine learning for materials discovery will be discussed, as well as how open source packages were leveraged to achieve results faster and more efficiently.
\end{abstract}


\section{Introduction}

\begin{figure}
  \includegraphics[width=\linewidth]{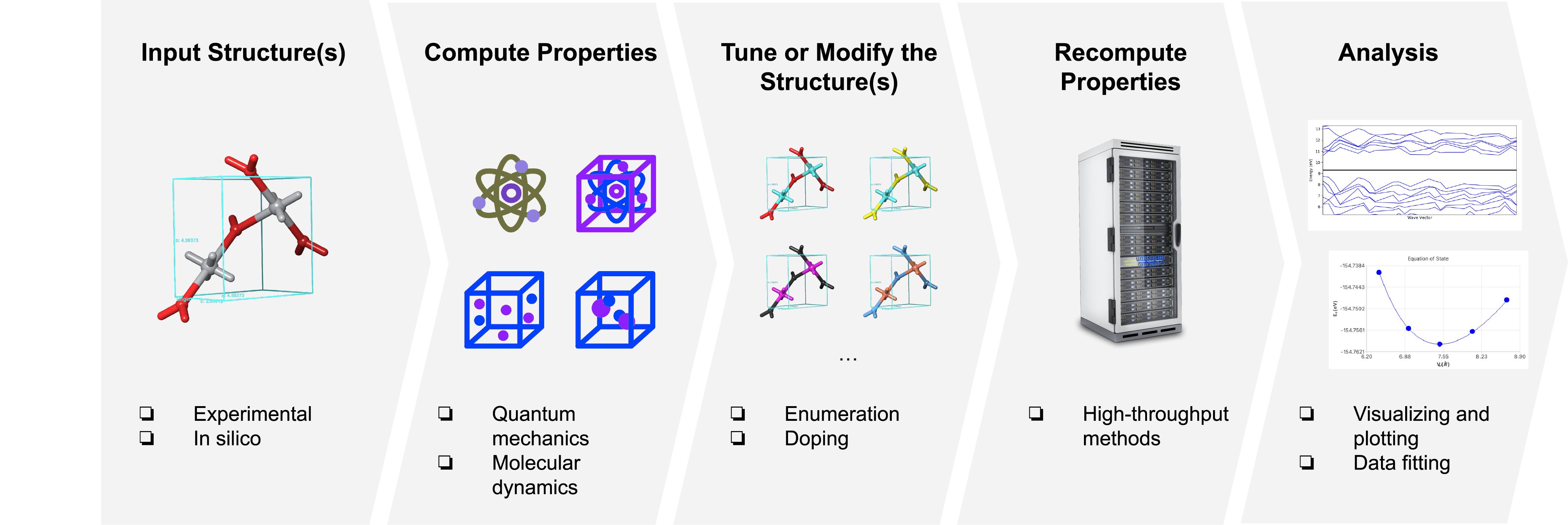}
  \caption{Example of a workflow for computational materials discovery.}
  \label{fig:fig1}
\end{figure}

A common materials discovery practice or workflow is to start with reading an experimental structure of a material or generating a structure in silico, computing its properties of interest (e.g. elastic constants, electrical conductivity), tuning the material by modifying its structure (e.g. doping) or adding and removing atoms (deposition, evaporation), and then recomputing the properties of the modified material (Figure \ref{fig:fig1}).
Computational materials discovery leverages such workflows to empower researchers to explore vast design spaces and uncover root causes without (or in conjunction with) laboratory experimentation.

Software tools for computational materials discovery can be facilitated by utilizing existing libraries that cover the fundamental mathematics used in the calculations in an optimized fashion. This use of existing libraries allows developers to devote more time to developing new features instead of re-inventing established methods. As a result, such a complementary approach improves the performance of computational materials software and reduces overall maintenance.

The Schr\"odinger Materials Science Suite \cite{Schr} is a proprietary computational chemistry/physics platform that streamlines materials discovery workflows into a single graphical user interface (Materials Science Maestro). The interface is a single portal for structure building and enumeration, physics-based modeling and machine learning, visualization and analysis. Tying together the various modules are a wide variety of scientific packages, some of which are proprietary to Schr\"odinger, Inc., some of which are open-source and many of which blend the two to optimize capabilities and efficiency. For example, the main simulation engine for molecular quantum mechanics is the Jaguar \cite{Jaguar} proprietary code. The proprietary classical molecular dynamics code Desmond (distributed by Schr\"odinger, Inc.) \cite{Desmond} is used to obtain physical properties of soft materials, surfaces and polymers. For periodic quantum mechanics, the main simulation engine is the open source code Quantum ESPRESSO (QE) \cite{QE}. One of the co-authors of this proceedings (A. Fonari) contributes to the QE code in order to make integration with the Materials Suite more seamless and less error-prone. As part of this integration, support for using the portable \texttt{XML} format for input and output in QE has been implemented in the open source Python package qeschema \cite{qeschema}.

\begin{figure}
  \includegraphics[width=\linewidth]{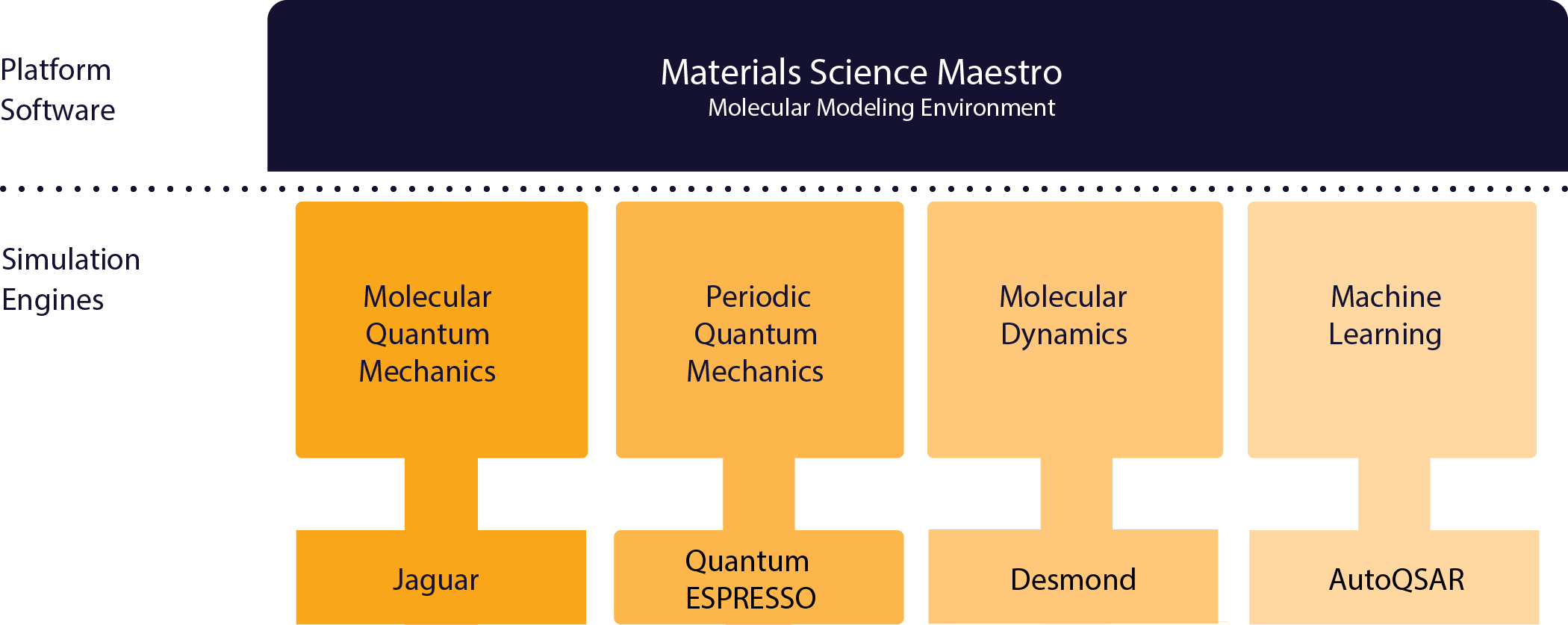}
  \caption{Some example products that compose the Schr\"odinger Materials Science Suite.}
  \label{fig:fig2}
\end{figure}

Figure \ref{fig:fig2} gives an overview of some of the various products that compose the Schr\"odinger Materials Science Suite. The various workflows are implemented mainly in Python (some of them described below), calling on proprietary or open-source code where appropriate, to improve the performance of the software and reduce overall maintenance.

The materials discovery cycle can be run in a high-throughput manner, enumerating different structure modifications in a systematic fashion, such as doping ratio in a semiconductor or depositing different adsorbates. As we will detail herein, there are several open source packages that allow the user to generate a large number of structures, run calculations in high throughput manner and analyze the results. For example, the open source package pymatgen \cite{pymatgen} facilitates generation and analysis of periodic structures. It can generate inputs for and read outputs of QE, the commercial codes VASP and Gaussian, and several other formats. To run and manage workflow jobs in a high-throughput manner, open source packages such as Custodian \cite{pymatgen} and AiiDA \cite{AiiDA} can be used.

\section{Materials import and generation}

For reading and writing of material structures, several open source packages (e.g. OpenBabel \cite{Obabel}, RDKit \cite{RDKit}) have implemented functionality for working with several commonly used formats (e.g. CIF, PDB, mol, xyz). Periodic structures of materials, mainly coming from single crystal X-ray/neutron diffraction experiments, are distributed in CIF (Crystallographic Information File), PDB (Protein Data Bank) and lately mmCIF formats \cite{Formats}. Correctly reading experimental structures is of significant importance, since the rest of the materials discovery workflow depends on it. In addition to atom coordinates and periodic cell information, structural data also contains symmetry operations (listed explicitly or by the means of providing a space group) that can be used to decrease the number of computations required for a particular system by accounting for symmetry. This can be important, especially when scaling high-throughput calculations. From file, structure is read in a structure object through which atomic coordinates (as a NumPy array) and chemical information of the material can be accessed and updated. Structure object is similar to the one implemented in open source packages such as pymatgen \cite{pymatgen} and ASE \cite{ASE}. All the structure manipulations during the workflows are done by using structure object interface (see structure deformation example below).Example of Structure object definition in pymatgen:
\begin{lstlisting}[language=Python]
    class Structure:

        def __init__(self, lattice, species, coords, ...):
            """Create a periodic structure."""
\end{lstlisting}

One consideration of note is that PDB, CIF and mmCIF structure formats allow description of the positional disorder (for example, a solvent molecule without a stable position within the cell which can be described by multiple sets of coordinates). Another complication is that experimental data spans an interval of almost a century: one of the oldest crystal structures deposited in the Cambridge Structural Database (CSD) \cite{CSD} dates to 1924 \cite{Grph}. These nuances and others present nontrivial technical challenges for developers. Thus, it has been a continuous effort by Schr\"odinger, Inc. (at least 39 commits and several weeks of work went into this project) and others to correctly read and convert periodic structures in OpenBabel. By version 3.1.1 (the most recent at writing time), the authors are not aware of any structures read incorrectly by OpenBabel. In general, non-periodic molecular formats are simpler to handle because they only contain atom coordinates but no cell or symmetry information. OpenBabel has Python bindings but due to the GPL license limitation, it is called as a subprocess from the Schr\"odinger Materials Suite.

Another important consideration in structure generation is modeling of substitutional disorder in solid alloys and materials with point defects (intermetallics, semiconductors, oxides and their crystalline surfaces). In such cases, the unit cell and atomic sites of the crystal or surface slab are well defined while the chemical species occupying the site may vary. In order to simulate substitutional disorder, one must generate the ensemble of structures that includes all statistically significant atomic distributions in a given unit cell. This can be achieved by a brute force enumeration of all symmetrically unique atomic structures with a given number of vacancies, impurities or solute atoms. The open source library enumlib \cite{Enumlib} implements algorithms for such a systematic enumeration of periodic structures. The enumlib package consists of several Fortran binaries and Python scripts that can be run as a subprocess (no Python bindings). This allows the user to generate a large set of symmetrically nonequivalent materials with different compositions (e.g. doping or defect concentration).

Recently, we applied this approach in simultaneous study of the activity and stability of Pt based core-shell type catalysts for the oxygen reduction reaction \cite{TM}. We generated a set of stable doped Pt/transition metal/nitrogen surfaces using periodic enumeration. Using QE to perform periodic density functional theory (DFT) calculations, we assessed surface phase diagrams for Pt alloys and identified the avenues for stabilizing the cost effective core-shell systems by a judicious choice of the catalyst core material. Such catalysts may prove critical in electrocatalysis for fuel cell applications.

\section{Workflow capabilities}

\begin{figure}
  \includegraphics[width=\linewidth]{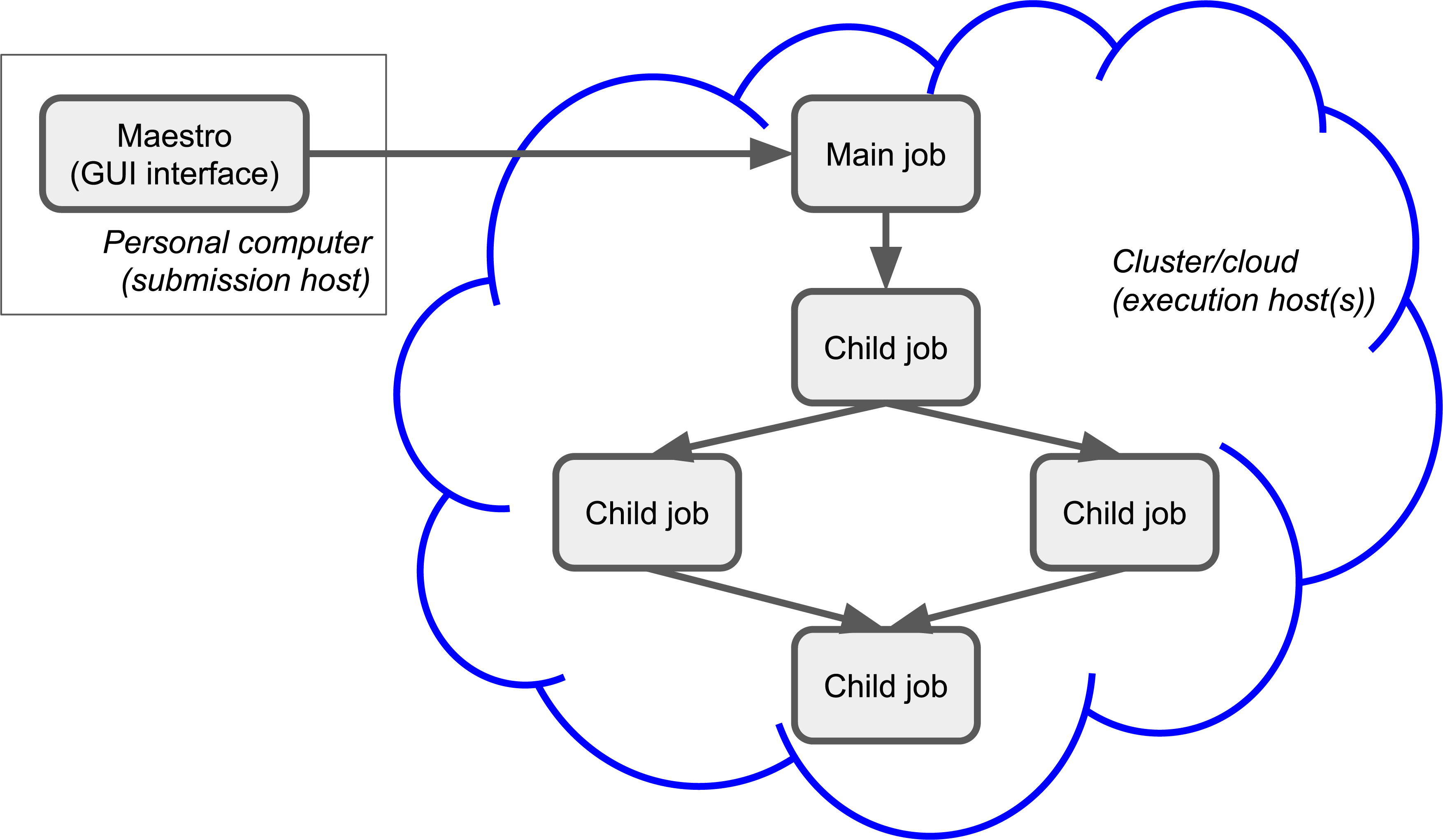}
  \caption{Example of the job submission process.}
  \label{fig:fig3}
\end{figure}

In the last section, we briefly described a complete workflow from structure generation and enumeration to periodic DFT calculations to analysis. In order to be able to run a massively parallel screening of materials, a highly scalable and stable queuing system (job scheduler) is required. We have implemented a job queuing system on top of the most used queuing systems (LSF, PBS, SGE, SLURM, TORQUE, UGE) and exposed a Python API to submit and monitor jobs. In line with technological advancements, cloud is also supported by means of a virtual cluster configured with SLURM. This allows the user to submit a large number of jobs, limited only by SLURM scheduling capabilities and cloud resources. In order to accommodate job dependencies in workflows, for each job, a parent job (or multiple parent jobs) can be defined forming a directed graph of jobs (Figure \ref{fig:fig3}).

There could be several reasons for a job to fail. Depending on the reason of failure, there are several restart and recovery mechanisms in place. The lowest level is the restart mechanism (in SLURM it is called \texttt{requeue}) which is performed by the queuing system itself. This is triggered when a node goes down. On the cloud, preemptible instances (nodes) can go offline at any moment. In addition, workflows implemented in the proprietary Schr\"odinger Materials Science Suite have built-in methods for handling various types of failure. For example, if the simulation is not converging to a requested energy accuracy, it is wasteful to blindly restart the calculation without changing some input parameters. However, in the case of a failure due to full disk space, it is reasonable to try restart with hopes to get a node with more empty disk space. If a job fails (and cannot be restarted), all its children (if any) will not start, thus saving queuing and computational time.

Having developed robust systems for running calculations, job queuing and troubleshooting (autonomously, when applicable), the developed workflows have allowed us and our customers to perform massive screenings of materials and their properties. For example, we reported a massive screening of 250,000 charge-conducting organic materials, totaling approximately 3,619,000 DFT SCF (self-consistent field) single-molecule calculations using Jaguar that took 457,265 CPU hours ($\approx$52 years) \cite{CScreen}. Another similar case study is the high-throughput molecular dynamics simulations (MD) of thermophysical properties of polymers for various applications \cite{MDS}. There, using Desmond we computed the glass transition temperature ($T_g$) of 315 polymers and compared the results with experimental measurements \cite{Bicerano}. This study took advantage of GPU (graphics processing unit) support as implemented in Desmond, as well as the job scheduler API described above.

Other workflows implemented in the Schr\"odinger Materials Science Suite utilize open source packages as well. For soft materials (polymers, organic small molecules and substrates composed of soft molecules), convex hull and related mathematical methods are important for finding possible accessible solvent voids (during submerging or sorption) and adsorbate sites (during molecular deposition). These methods are conveniently implemented in the open source SciPy \cite{Scipy} and NumPy \cite{Numpy} packages. Thus, we implemented molecular deposition and evaporation workflows by using the Desmond MD engine as the backend in tandem with the convex hull functionality. This workflow enables simulation of the deposition and evaporation of the small molecules on a substrate. We utilized the aforementioned deposition workflow in the study of organic light-emitting diodes (OLEDs), which are fabricated using a stepwise process, where new layers are deposited on top of previous layers. Both vacuum and solution deposition processes have been used to prepare these films, primarily as amorphous thin film active layers lacking long-range order. Each of these deposition techniques introduces changes to the film structure and consequently, different charge-transfer and luminescent properties \cite{Deposition}.

As can be seen from above, a workflow is usually some sort of structure modification through the structure object with a subsequent call to a backend code and analysis of its output if it succeeds. Input for the next iteration depends on the output of the previous iteration in some workflows. Due to the large chemical and manipulation space of the materials, sometimes it very tricky to keep code for all workflows follow the same code logic. For every workflow and/or functionality in the Materials Science Suite, some sort of peer reviewed material (publication, conference presentation) is created where implemented algorithms are described to facilitate reproducibility.

\section{Data fitting algorithms and use cases}

Materials simulation engines for QM, periodic DFT, and classical MD (referred to herein as backends) are frequently written in compiled languages with enabled parallelization for CPU or GPU hardware. These backends are called from Python workflows using the job queuing systems described above. Meanwhile, packages such as SciPy and NumPy provide sophisticated numerical function optimization and fitting capabilities. Here, we describe examples of how the Schr\"odinger suite can be used to combine materials simulations with popular optimization routines in the SciPy ecosystem.

\begin{figure}
  \includegraphics[width=\linewidth]{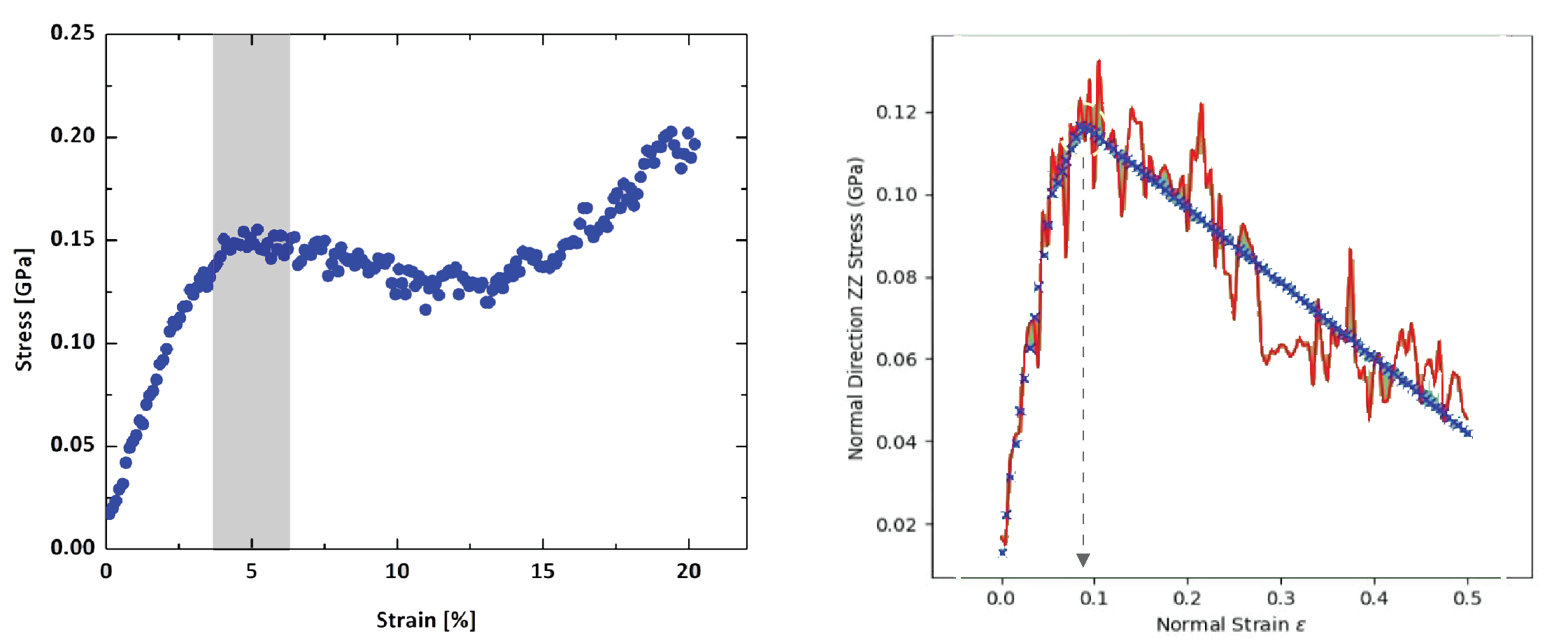}
  \caption{Left: The uniaxial stress/strain curve of a polymer calculated using Desmond through the stress strain workflow. The dark grey band indicates an inflection that marks the yield point. Right: Constant strain simulation with convex analysis indicates elongation at yield. The red curve shows simulated stress versus strain. The blue curve shows convex analysis.}
  \label{fig:fig4}
\end{figure}

Recently we implemented convex analysis of the stress strain curve (as described here \cite{Patrone}). \texttt{scipy.optimize.minimize} is used for a constrained minimization with boundary conditions of a function related to the stress strain curve. The stress strain curve is obtained from a series of MD simulations on deformed cells (cell deformations are defined by strain type and deformation step). The pressure tensor of a deformed cell is related to stress. This analysis allowed prediction of elongation at yield for high density polyethylene polymer. Figure \ref{fig:fig4} shows obtained calculated yield of 10\% vs. experimental value within 9-18\% range \cite{Convex}.

The \texttt{scipy.optimize} package is used for a least-squares fit of the bulk energies at different cell volumes (compressed and expanded) in order to obtain the bulk modulus and equation of state (EOS) of a material. In the Schr\"odinger suite this was implemented as a part of an EOS workflow, in which fitting is performed on the results obtained from a series of QE calculations performed on the original as well as compressed and expanded (deformed) cells. An example of deformation applied to a structure in pymatgen:

\begin{lstlisting}[language=Python]
   from pymatgen.analysis.elasticity import strain
   from pymatgen.core import lattice
   from pymatgen.core import structure

   deform = strain.Deformation([
      [1.0, 0.02, 0.02],
      [0.0, 1.0, 0.0],
      [0.0, 0.0, 1.0]])

   latt = lattice.Lattice([
      [3.84, 0.00, 0.00],
      [1.92, 3.326, 0.00],
      [0.00, -2.22, 3.14],
   ])

   st = structure.Structure(
      latt,
      ["Si", "Si"],
      [[0, 0, 0], [0.75, 0.5, 0.75]])

   strained_st = deform.apply_to_structure(st)
\end{lstlisting}

This is also an example of loosely coupled (embarrassingly parallel) jobs. In particular, calculations of the deformed cells only depend on the bulk calculation and do not depend on each other. Thus, all the deformation jobs can be submitted in parallel, facilitating high-throughput runs.

Structure refinement from powder diffraction experiment is another example where more complex optimization is used. Powder diffraction is a widely used method in drug discovery to assess purity of the material and discover known or unknown crystal polymorphs \cite{Powder}. In particular, there is interest in fitting of the experimental powder diffraction intensity peaks to the indexed peaks (Pawley refinement) \cite{Jansen}. Here we employed the open source lmfit package \cite{Lmfit} to perform a minimization of the multivariable Voigt-like function that represents the entire diffraction spectrum. This allows the user to refine (optimize) unit cell parameters coming from the indexing data and as the result, goodness of fit ($R$-factor) between experimental and simulated spectrum is minimized.

\section{Machine learning techniques}

Of late, there is great interest in machine learning assisted materials discovery. There are several components required to perform machine learning assisted materials discovery.
In order to train a model, benchmark data from simulation and/or experimental data is required. Besides benchmark data, computation of the relevant descriptors is required (see below). Finally, a model based on benchmark data and descriptors is generated that allows prediction of properties for novel materials. There are several techniques to generate the model, such as linear or non-linear fitting to neural networks. Tools include the open source DeepChem \cite{DeepChem} and AutoQSAR \cite{AutoQSAR} from the Schr\"odinger suite. Depending on the type of materials, benchmark data can be obtained using different codes available in the Schr\"odinger suite:

\begin{itemize}
    \item small molecules and finite systems - Jaguar
    \item periodic systems - Quantum ESPRESSO
    \item larger polymeric and similar systems - Desmond
\end{itemize}

Different materials systems require different descriptors for featurization. For example, for crystalline periodic systems, we have implemented several sets of tailored descriptors. Generation of these descriptors again uses a mix of open source and Schr\"odinger proprietary tools. Specifically:

\begin{itemize}
    \item elemental features such as atomic weight, number of valence electrons in *s*, *p* and *d*-shells, and electronegativity
    \item structural features such as density, volume per atom, and packing fraction descriptors implemented in the open source matminer package \cite{Matminer}
    \item intercalation descriptors such as cation and anion counts, crystal packing fraction, and average neighbor ionicity \cite{Sendek} implemented in the Schr\"odinger suite
    \item three-dimensional smooth overlap of atomic positions (SOAP) descriptors implemented in the open source DScribe package \cite{DScribe}.
\end{itemize}

We are currently training models that use these descriptors to predict properties, such as bulk modulus, of a set of Li-containing battery related compounds \cite{Chandrasekaran}. Several models will be compared, such as kernel regression methods (as implemented in the open source scikit-learn code \cite{SkLearn}) and AutoQSAR.

For isolated small molecules and extended non-periodic systems, RDKit can be used to generate a large number of atomic and molecular descriptors. A lot of effort has been devoted to ensure that RDKit can be used on a wide variety of materials that are supported by the Schr\"odinger suite. At the time of writing, the 4th most active contributor to RDKit is Ricardo Rodriguez-Schmidt from Schr\"odinger \cite{RDKitC}.

\begin{figure}
  \includegraphics[width=\linewidth]{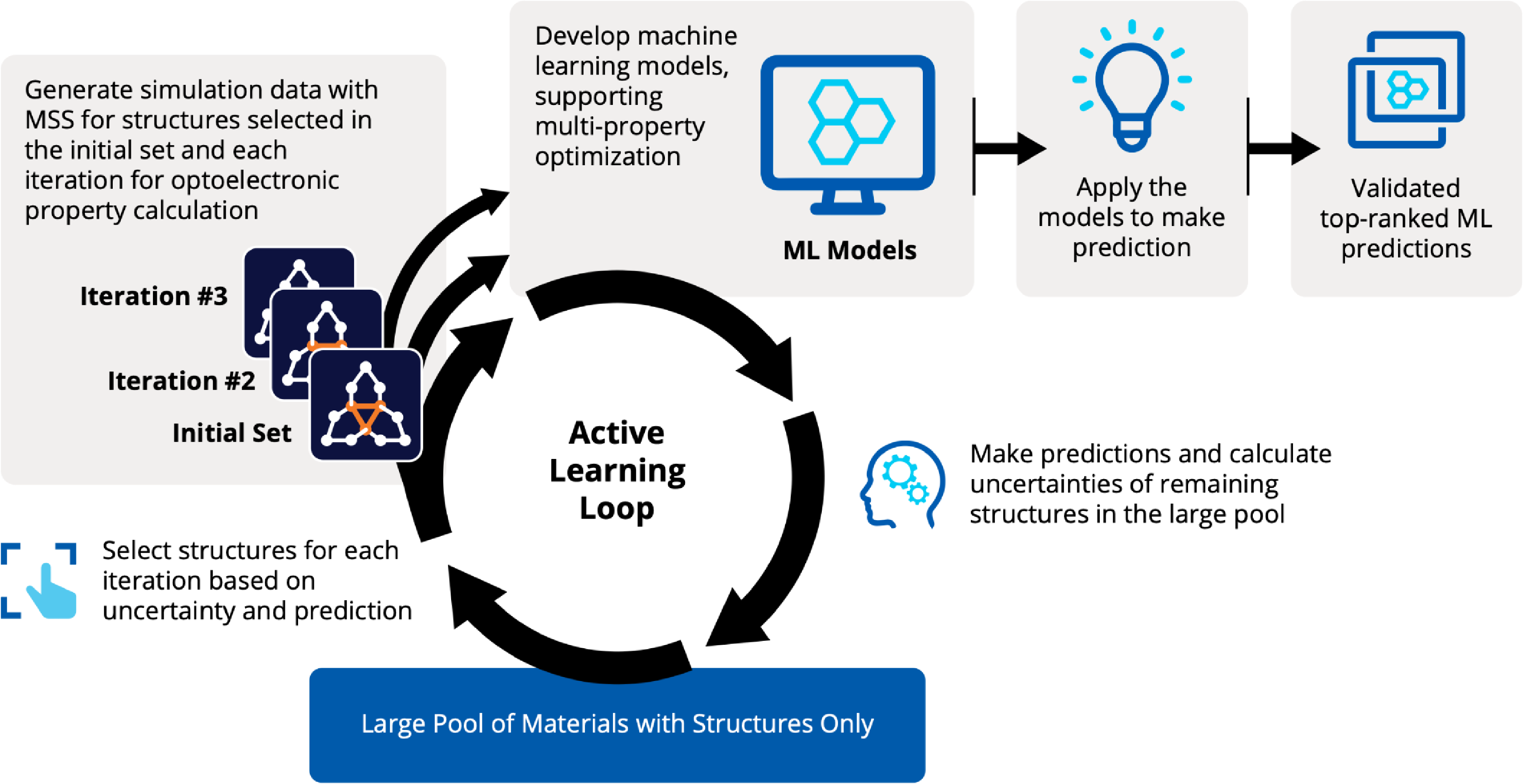}
  \caption{Active learning workflow for the design and discovery of novel optoelectronics molecules.}
  \label{fig:fig5}
\end{figure}

Recently, active learning (AL) combined with DFT has received much attention to address the challenge of leveraging exhaustive libraries in materials informatics \cite{Vasudevan}, \cite{Schleder}. On our side, we have implemented a workflow that employs active learning (AL) for intelligent and iterative identification of promising materials candidates within a large dataset. In the framework of AL, the predicted value with associated uncertainty is considered to decide what materials to be added in each iteration, aiming to improve the model performance in the next iteration (Figure \ref{fig:fig5}).

\begin{figure}
  \includegraphics[width=\linewidth]{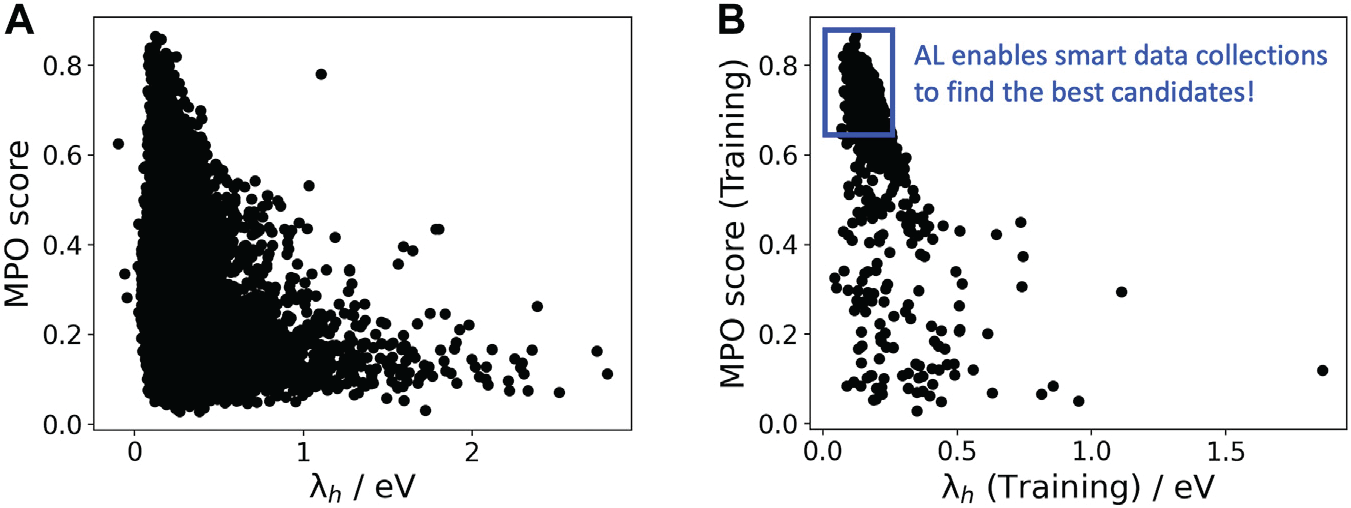}
  \caption{A: MPO score of all materials in the HTL dataset. B: Those used in the training set as a function of the hole reorganization energy ($\lambda_h$).}
  \label{fig:fig6}
\end{figure}

Since it could be important to consider multiple properties simultaneously in material discovery, multiple property optimization (MPO) has also been implemented as a part of the AL workflow \cite{Kwak}. MPO allows scaling and combining multiple properties into a single score. We employed the AL workflow to determine the top candidates for hole (positively charged carrier) transport layer (HTL) by evaluating 550 molecules in 10 iterations using DFT calculations for a dataset of $\approx$9,000 molecules \cite{Abroshan}. Resulting model was validated by randomly picking a molecule from the dataset, computing properties with DFT and comparing those to the predicted values. According to the semiclassical Marcus equation \cite{Marcus}, high rates of hole transfer are inversely proportional to hole reorganization energies. Thus, MPO scores were computed based on minimizing hole reorganization energy and targeting oxidation potential to an appropriate level to ensure a low energy barrier for hole injection from the anode into the emissive layer. In this workflow, we used RDKit to compute descriptors for the chemical structures.
These descriptors generated on the initial subset of structures are given as vectors to an algorithm based on Random Forest Regressor as implemented in scikit-learn. Bayesian optimization is employed to tune the hyperparameters of the model. In each iteration, a trained model is applied for making predictions on the remaining materials in the dataset. Figure \ref{fig:fig6} (A) displays MPO scores for the HTL dataset estimated by AL as a function of hole reorganization energies that are separately calculated for all the materials. This figure indicates that there are many materials in the dataset with desired low hole reorganization energies but are not suitable for HTL due to their improper oxidation potentials, suggesting that MPO is important to evaluate the optoelectronic performance of the materials. Figure \ref{fig:fig6} (B) presents MPO scores of the materials used in the training dataset of AL, demonstrating that the feedback loop in the AL workflow efficiently guides the data collection as the size of the training set increases.

To appreciate the computational efficiency of such an approach, it is worth noting that performing DFT calculations for all of the 9,000 molecules in the dataset would increase the computational cost by a factor of 15 versus the AL workflow. It seems that AL approach can be useful in the cases where problem space is broad (like chemical space), but there are many clusters of similar items (similar molecules). In this case, benchmark data is only needed for few representatives of each cluster. We are currently working on applying this approach to train models for predicting physical properties of soft materials (polymers).

\section{Conclusions}

We present several examples of how Schr\"odinger Materials Suite integrates open source software packages. There is a wide range of applications in materials science that can benefit from already existing open source code. Where possible, we report issues to the package authors and submit improvements and bug fixes in the form of the pull requests. We are thankful to all who have contributed to open source libraries, and have made it possible for us to develop a platform for accelerating innovation in materials and drug discovery. We will continue contributing to these projects and we hope to further give back to the scientific community by facilitating research in both academia and industry. We hope that this report will inspire other scientific companies to give back to the open source community in order to improve the computational materials field and make science more reproducible.

\section{Acknowledgments}

The authors acknowledge Bradley Dice and Wenduo Zhou for their valuable comments during the review of the manuscript.

\printbibliography
\end{document}